\documentclass[a4paper,11pt]{article}
\usepackage{pos}
\usepackage{physics}
\usepackage{subcaption}
\usepackage{todonotes}
\usepackage{amsthm}
\usepackage{amsfonts}
\usepackage{bbold}
\usepackage{hyperref}
\usepackage{setspace}
\title{Charm physics with a tmQCD mixed action}

\author[c]{A. Bussone}
\author*[a,c]{A. Conigli}
\author[e]{J. Frison}
\author[a,b]{G. Herdo\'iza}
\author[a,b]{C. Pena}
\author[d]{D. Preti} 
\author[a,b]{J.\'A. Romero}
\author[a,b]{J. Ugarrio}
\affiliation[a]{Instituto de F\'isica Te\'orica UAM-CSIC, c/ Nicol\'as Cabrera 13-15, Universidad Aut\'onoma de Madrid, E-28049 Madrid, Spain}

\affiliation[b]{Department of Theoretical Physics, Universidad Aut\'onoma de Madrid, E-28049 Madrid, Spain}

\affiliation[c]{Institut f\"{u}r Physik, Humboldt Universit\"{a}t zu Berlin, Newtonstra{\ss}e 15, 12489 Berlin} 

\affiliation[d]{INFN, Sezione di Torino Via Pietro Giuria 1, I-10125 Turin, Italy}

\affiliation[e]{John von Neumann Institute for Computing, DESY, Platanenallee 6, 15738 Zeuthen,
	Germany}

\emailAdd{alessandro.conigli@uam.es}

\abstract{We report on our ongoing determination of the charm quark mass and the masses and decay constants of various charmed mesons, obtained within a mixed-action setup. We employ $N_f=2+1$ CLS ensembles combined with a Wilson twisted mass valence action that eliminates the leading $O(a)$ effects from our target observables. Alongside our preliminary results, we will discuss an exploration of GEVP techniques aimed at optimizing the precision in view of the extension of the computation to heavier quark masses. We study the chiral-continuum extrapolation of decay constants for charm quark observables and the renormalized charm quark mass. }

\FullConference{%
 The 38th International Symposium on Lattice Field Theory, LATTICE2021
  26th-30th July, 2021
  Zoom/Gather@Massachusetts Institute of Technology
}

\begin{document}

\maketitle

\section{Introduction}
\noindent
Lattice QCD (LQCD) simulations allow to  quantitatively study the realm of flavour physics, one of the most promising sectors for the search of physics beyond the Standard Model (SM). In particular, processes involving heavy quarks are essential to achieve reliable estimates of SM parameters and to examine the compatibility between the theory predictions and the experimental measurements. Our main goal is to compute hadronic matrix elements in the charm sector necessary to the determination of quantities of  phenomenological interest such as  the CKM matrix elements, reaching precisions that allow to improve the state-of-the-art.
\\
In order to achieve reliable estimates  we have developed a setup that aims at controlling the systematic uncertainties in LQCD simulations involving charm quarks \cite{tmQCD}. In particular, we exploit a mixed action approach where in the sea sector we employ CLS  $N_f= 2+1$   gauge ensembles \cite{cls1,cls2} with open boundary conditions in the time direction and four fine values of the lattice spacing. \\In the valence sector we take advantage of a twisted mass action with $N_f= 2+1+1$ flavours at  full twist. This ensures automatic $O(a)$ improvement of the observables up to residual lattice artifacts arising from the sea light quark masses \cite{Bussone:2018ljj}.
\\
For charmed semileptonics results from a tmQCD mixed action we refers to \cite{julien2021}. 

We present an update from \cite{Ugarrio:2018ghf} for the quark mass and for meson masses and decay constants  in the charm-quark sector, with a particular focus on the chiral-continuum extrapolations to the physical point.

\section{Lattice setup}

\subsection{Sea sector} \label{sec:mass_shift}
\noindent
In this work we use  the gauge configurations  produced within the CLS initiative \cite{cls1}, whose ensembles adopt the tree-level improved L\"{u}scher-Weisz gauge action \cite{gauge_action}.
On the fermionic sector the action involves a Wilson Dirac operator for $N_f = 2+1$ flavours \cite{fermion_action} with the inclusion of the Sheikholeslami-Wolhert term to achieve non-perturbative $O(a)$ improvement \cite{csw}.

This work is performed with four values of the lattice spacing ranging from 0.087 fm  down to 0.050 fm.
In Table \ref{tab:ens}
we report the  list of CLS $N_f=2+1$ ensembles we used.
In particular these ensembles lie  along an approximate line of constant trace of the bare quark mass matrix,
\begin{equation}
	\tr(M_q) = 2m_{q,u} + m_{q,s} = \text{const},
\end{equation}
where $m_{q,f} = m_{0,f} - m_{\text{cr}}$.
In practice, deviations from $\tr(M_q)=\text{const}$ have been observed to be large at the coarsest lattice spacing \cite{Bruno:2016plf}. Therefore, in the approach to the physical point we redefine the chiral trajectory by imposing 
\begin{equation}
	\phi_4 \equiv 8t_0 \bigg(
	\frac{1}{2}m_\pi^2 + m_K^2
	\bigg) = \phi_4^{\text{phys}} \qquad \phi_2 = 8t_0 m_\pi^2.
\end{equation}
where the gluonic quantity $t_0$ is defined from  the Wilson flow \cite{Wilflow}. The  dependencies of the observables on light sea quark masses are parametrised by $\phi_2$.

To ensures that $\phi_4$ passes across the physical point it is essential to apply small mass corrections to the bare quark masses  through a low order Taylor expansion, as explained in \cite{Bruno:2016plf}. Given a generic derived observable $\mathcal{O}$, its derivative with respect to the sea quark mass is given by 
\begin{equation}
	\dv{\langle\mathcal{O}\rangle}{m_{q,i}} =
	\bigg\langle \pdv{\mathcal{O}}{m_{q,i}} \bigg\rangle
	-
	\bigg{\langle}
	\mathcal{O}\pdv{S}{m_{q,i}} \bigg{\rangle}
	+
	\langle\mathcal{O}\rangle
	\bigg{\langle}
	\pdv{S}{m_{q,i}}\bigg{\rangle},
	\qquad \quad
	\langle\mathcal{O}\rangle \rightarrow
	\langle\mathcal{O}\rangle + \sum_{i=1}^{N_f}\Delta m_{q,i}\dv{\langle\mathcal{O}\rangle}{m_{q,i}},
\end{equation}
where $S$ is the action. The shift of the observable depending on sea quark flavours is then given by right-most expression in the above equation,
where the value of $\Delta m_{q,i}$ is obtained on each ensemble with an iterative procedure that matches $\phi_4$ to its physical value. The value of $\Delta m_{q,i}$ is chosen to be the same for all quark flavours.

\begin{table}[h!]
	\begin{center}	
		\begin{tabular}{c  c  c  c  c  c  c}
			\hline
			\text{Id} & $\beta $ & $N_s$ & $N_t$ & $m_\pi$ [MeV] & $m_K$ [MeV] & $M_\pi L $\\
			\hline
			H101 & 3.4 & 32 & 96 & 420 & 420 & 5.8  \\
			H102 & 3.4  & 32 & 96 & 350 & 440 & 5.9 \\
			\hline 
			H400 & 3.46 & 32 & 96 & 420 & 420 & 5.2  \\
			\hline
			N200 & 3.55 & 48 & 128 & 280 & 460 & 4.4  \\
			N202 & 3.55 & 48 & 128 & 420 & 420 & 6.5  \\
			N203 & 3.55 & 48 & 128 & 340 & 440 & 5.4  \\
			\hline
			N300 & 3.70 & 48 & 128 & 420 & 420 & 5.1 \\
			J303 & 3.70 & 64 & 196 & 260 & 470 & 4.1  \\
			\hline
		\end{tabular}
		\caption{List of CLS $N_f=2+1$ ensembles used in the present study \cite{cls1}. $N_s$ and $N_t$ refer to the spatial and temporal extent of the lattice. Approximate values of the pion and kaon masses are provided. }
		\label{tab:ens}
	\end{center}
\end{table}

\subsection{Valence sector}
\noindent
The mixed action setup employs in the valence sector a Wilson twisted mass Dirac operator \cite{tm1} with $N_f=2+1+1$ valence flavours. This chiral rotation of the fields yields to an action in the twisted basis. In addition, we include a clover term and we  use the non-perturbative determination of the Sheikholeslami-Wohlert coefficient for $N_f=3$ \cite{csw_np},
\begin{equation}
	D_{\text{tm}} = \frac{1}{2}\sum_{\mu=0}^{3}\big[
	\gamma_\mu ( \nabla^*_\mu + \nabla_\mu) - a \nabla^*_\mu \nabla_\mu
	\big]
	+
	\frac{i}{4}a c_{\rm{\scriptscriptstyle SW}}\sum_{\mu,\nu=0}^{3} \sigma_{\mu\nu} \hat{F}_{\mu\nu} + \mathbf{m}_0+ i\gamma^5\boldsymbol{\mu}_0,
\end{equation}
where $\nabla_\mu$ and $\nabla_\mu^*$ are the forward and backward covariant derivatives.
The maximal twist regime is achieved by imposing the bare standard mass matrix to be equal to the critical mass $ \mathbf{m_0} = \mathbb{1} \mathbf{m_{cr}} $. In this setup the twisted mass matrix $\boldsymbol{\mu}_0 = \text{diag}(+\mu_{0,l},- \mu_{0,l},- \mu_{0,s},+ \mu_{0,c})$ is associated to the physical bare quark masses.

In order to set the bare quark mass $\boldsymbol{m}_0$ to its critical value we tuned the valence PCAC quark mass to vanish in the light sector. This condition guarantees the full twist regime.

To recover the unitarity of the theory we perform the matching between sea and valence quark masses by imposing that the light pseudoscalar masses are equal in both sectors. More details on the matching procedure can be found in \cite{Bussone:2019mlt} and in the recent update \cite{gregorio}.

In our simulations the charm quark is partially quenched, hence the matching procedure for the charm sector requires a different strategy. To tackle the issue we simulate at three different values of the charm twisted mass $\mu_c$ and we  follow two different paths to tune the physical charm quark mass: we either employ the flavour averaged $D$ meson mass $m_{\bar{D}} = 2/3m_D +1/3m_{D_s}$ or the charmonium $\eta_c$ effective mass, neglecting the disconnected contributions. The interpolation to  physical values is performed jointly with a chiral-continuum extrapolations, as discussed  in section \ref{sec:chir_cont}.

\section{Observables}
\noindent
The open boundary conditions in the Euclidean time direction modify the spectrum of the theory in the neighbourhood of the boundaries. In order to address the boundary effects, the sources of the two-point functions are set in the bulk, precisely in the middle of the lattice at  $y_0 = T/2$. The two-point functions are then given by 
\begin{equation}
		f_P(x_0,y_0) = a^6\sum_{\vec{x},\vec{y}}
		\langle P(x) P(y)\rangle \qquad 
		f_{A_\mu}(x_0,y_0) =  a^6\sum_{\vec{x},\vec{y}}
		\langle A_\mu(x) P(y)\rangle,
\end{equation}
where $P$ and $A_\mu$ are the pseudoscalar density and the the improved axial current, respectively:
\begin{equation}
	P(x) = \bar{\psi}(x)\gamma_5\psi(x) \qquad A_\mu(x) = \bar{\psi}(x)\gamma_\mu\gamma_5\psi(x)+ac_A\partial_\mu P(x).
\end{equation}

In this work we rely on the generalized eigenvalue problem (GEVP) variational method to compute the  spectrum and  matrix elements of charmed mesons \cite{gevp}. The basic idea behind the method is to consider a matrix of correlators   $C(t)$ and solve the associated GEVP problem
\begin{equation}\label{eq:gevp}
	C(t) v_n(t,t_0) = \lambda_n(t,t_0) C(t_0)v_n(t,t_0) \qquad n=1,\ldots,N, \quad t>t_0,
\end{equation}
where $N$ is the correlators matrix dimension.

We choose to avoid mixing of parity-odd and parity-even operators in a unique GEVP system to properly isolate the  ground states. Having  this in mind, we solve  Eq. (\ref{eq:gevp}) for the two matrices of correlators
\begin{equation}\label{eq:matrix}
	C_{PP}(t) = \bigg[\begin{matrix}
		\langle P(t)P(0)\rangle & \langle P(t+\tau)P(0)\rangle
		\\
		\langle P(t)P(-\tau)\rangle & \langle P(t+\tau)P(-\tau)\rangle
	\end{matrix}\bigg]
	\qquad
	C_{VV}(t) = \bigg[\begin{matrix}
		\langle A_k(t)A_k(0)\rangle & \langle A_k(t+\tau)A_k(0)\rangle
		\\
		\langle A_k(t)A_k(-\tau)\rangle & \langle A_k(t+\tau)A_k(-\tau)\rangle
	\end{matrix}\bigg]
\end{equation}
where $C_{PP}(t)$ and $C_{VV}(t)$ denote the pseudoscalar and vector matrix of correlators, respectively. We tested several values of  $\tau$, the time shift of the correlators,  and we observed a mild dependence of the final results for small value of $\tau$. In what follows we set $\tau=3$. 

\begin{center}
	\begin{figure}[h]
		\begin{subfigure}{0.5\textwidth}
			\includegraphics[width=0.9\linewidth, height=5cm]{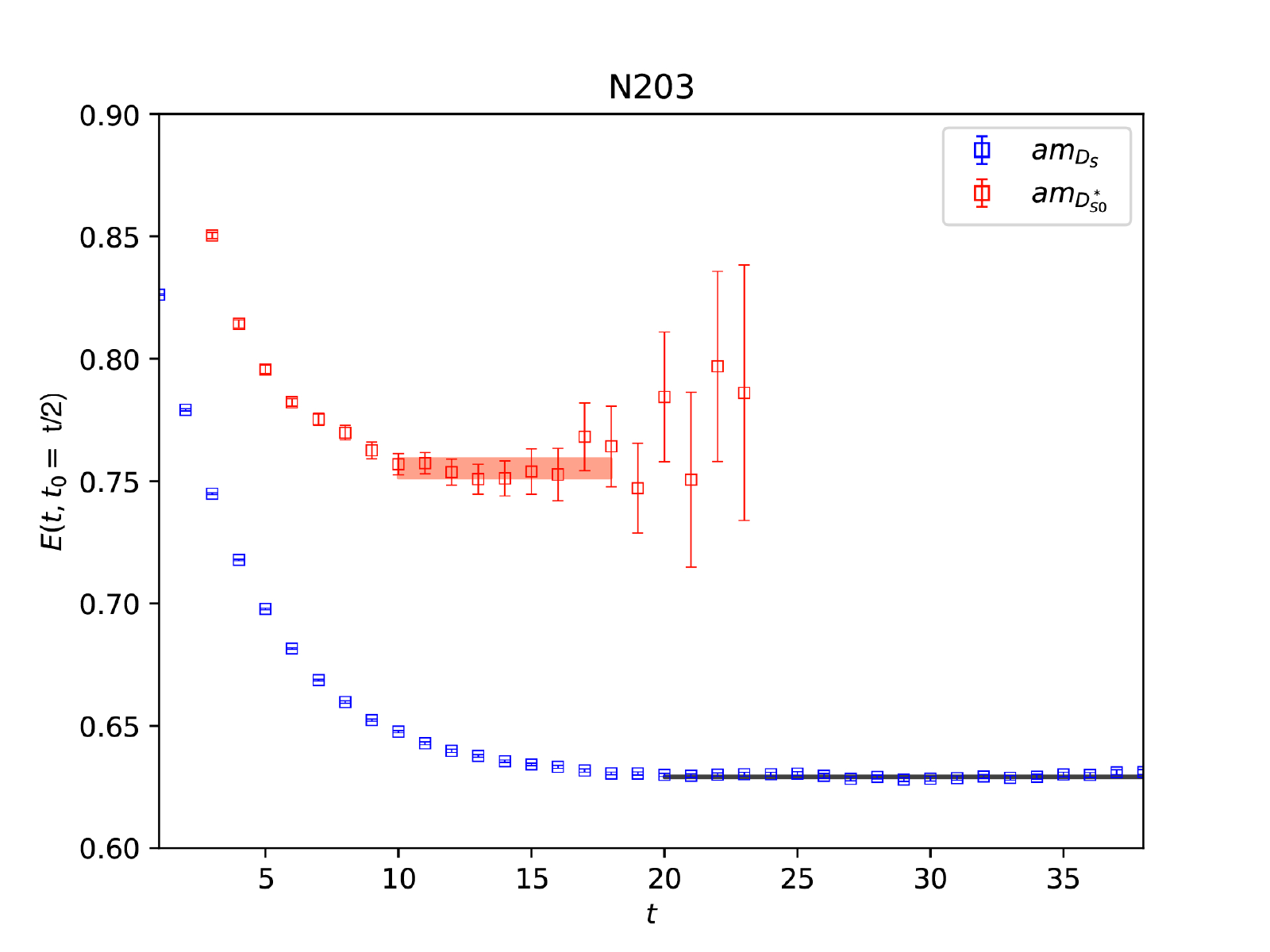} 
			\caption{Pseudoscalar  spectrum in the charm-light sector}
			\label{fig:subim1}
		\end{subfigure}
		\begin{subfigure}{0.5\textwidth}
			\includegraphics[width=0.9\linewidth, height=5cm]{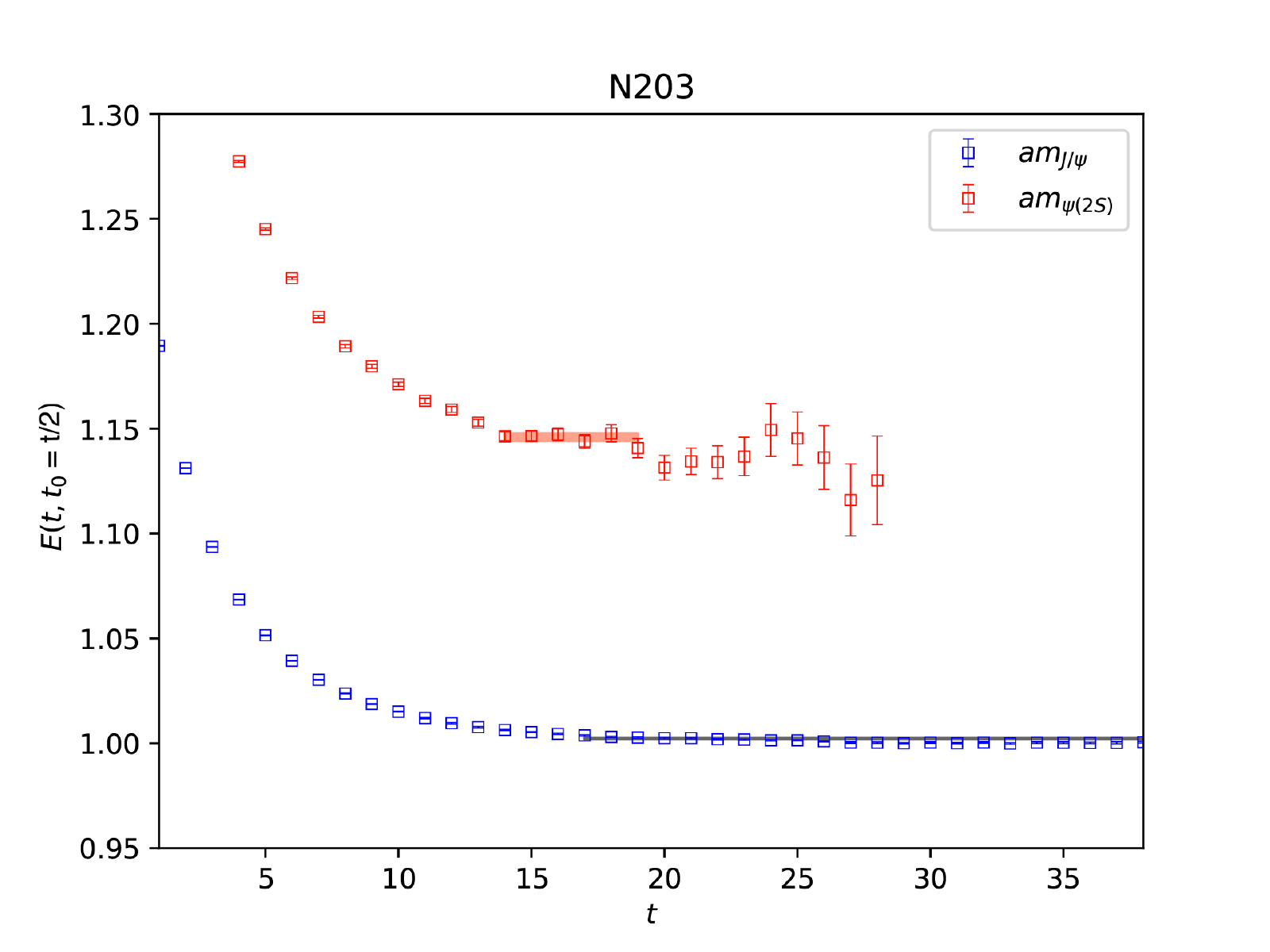}
			\caption{Vector spectrum in the charmonium sector}
			\label{fig:subim2}
		\end{subfigure}
		
		\caption{Meson masses of the ground state and first excited state as extracted from the GEVP. The shaded bands show the plateau regions as determined by the procedure explained in the text.}
		\label{fig:image2}
		
	\end{figure}
\end{center}

Once the GEVP is solved, we extract the energy spectrum via \cite{gevp}
\begin{equation}
	E_n^{\text{eff}} = \log(\frac{\lambda_n(t,t_0,\tau)}{\lambda_n(t+a, t_0,\tau)}) = E_n + p_1e^{-(E_{N+1}-E_n)t},
\end{equation} 
where $E_{N+1}$ is the first unresolved excited state.
The last term of the above equation is then used as a functional form to fit the effective energies, as described in detail in \cite{gevp}.
This allows to properly assess the systematic contribution from excited states  and eventually we define the plateau region by imposing the statistical error to be four times larger than the systematic one.\\ In Fig. \ref{fig:image2} we show the effective masses as extracted from the GEVP for two observables of interest in a representative ensemble. The same procedure is applied analogously  to the extraction of the matrix elements from the generalised eigenvectors.

\section{Chiral-continuum extrapolations} \label{sec:chir_cont}
\noindent
After having determined the decay constants, the meson spectrum and the renormalized charm quark mass from every gauge ensemble listed in Tab. \ref{tab:ens}, we can finally go ahead with a combined chiral-continuum extrapolation, essential to obtain observables at the physical point. The latter is defined by  $\phi_2= \phi_2^{\text{phys}}$ and $\phi_H= \phi_H^{\text{phys}}$ at zero lattice spacing, where 
\begin{equation}
	\phi_2 = 8t_0 m_\pi^2 \qquad \phi_H = \sqrt{8t_0} m_H \qquad m_H = m_{\bar{D}}, m_{\eta_c}
\end{equation}
are used to monitor the approach to the physical value of the bare light quark and the bare charm quark masses, respectively. Here $m_H$ labels the ground state of a charmed meson used to fix the charm quark mass, and we either employ the flavour averaged $D$ meson mass or the charmonium\footnote{We neglect all quark disconnected contributions.} $\eta_c$ mass.

For the lattice spacing dependence of the observables  we assume the leading cutoff effects to be of $O(a^2)$ as the mixed action at full twist ensures the absence of $O(a)$ cutoff effects\footnote{Apart from residual lattice artifacts proportional to the sea light quark masses. As explained in \cite{Bussone:2018ljj} these effects are negligible at the current precision. } and the relevant $O(a)$ improved renormalization constant are know non-perturbatively from \cite{ren_const}. Eventually our general ansatz for the lattice spacing dependence is parametrised by
\begin{equation}\label{eq:a_dep}
	c_\mathcal{O}(\phi_2, \phi_H, a) = \frac{a^2}{8t_0} \big(
	c_1 + c_2\phi_2 + c_3\phi_H^2
	\big)
	+
	\frac{a^4}{(8t_0)^2}\big(
	c_4 + c_5\phi_H^4
	\big).
\end{equation}
Here we  allow for cutoff terms describing the higher  $O(a^4)$ effects and we also consider cutoff effects proportional to the light quark masses. In the twisted mass formulation of LQCD at maximal twist, all the odd powers of the lattice spacing are suppressed. 

For the continuum mass dependence of the observables of interest we employ the functional form 
\begin{equation}\label{eq:cont_dep}
	\sqrt{8t_0}\mathcal{O}^{\text{cont}}(\phi_2, \phi_H,0) = p_0 + p_1\phi_2 + p_2\phi_H.
\end{equation}
We stress out that the light quark mass dependence is dominated  by the sea pion mass $\phi_2$ only, since the mass shift corrections to the chiral trajectory ensure the kaon masses to be fixed by the condition $\tr(M_q^R)=\text{const}$ as explained in Sec. \ref{sec:mass_shift}. More sophisticated ans\"{a}tze, \textit{e.g.} including chiral logarithms, will be studied as more ensembles are incorporated.

Finally to arrive at a combined model we follow a similar strategy as proposed in \cite{charm_quark} by either adding linearly or multiplying non-linearly Eqs. \ref{eq:a_dep} and \ref{eq:cont_dep}:
\begin{equation}\label{eq:tot_model}
	\begin{split}
		\sqrt{8t_0}\mathcal{O}^{\text{linear}}(\phi_2,\phi_H, a) &=
		\sqrt{8t_0}\mathcal{O}^{\text{cont}}(\phi_2,\phi_H,0) + c_\mathcal{O}(\phi_2,\phi_H,a),
	\\
	\sqrt{8t_0}\mathcal{O}^{\text{non-lin}}(\phi_2,\phi_H, a) &= 
	\sqrt{8t_0}\mathcal{O}^{\text{cont}}(\phi_2,\phi_H,0) \big(1+ c_\mathcal{O}(\phi_2,\phi_H,a)\big).	
\end{split}
\end{equation}
Then, by considering all the possible combinations of the coefficients $c_i$ in Eq. (\ref{eq:a_dep}), we end up with 64 different fit models. Since we are dealing with highly correlated data the uncorrelated $\chi^2$ does not yield reliable estimates of the fit paramaters. In practice we bypass this problem by employing the so-called $\chi^2$ expected, $\chi^2_{\text{exp}}$, as estimator for the goodness of a fit \cite{chiexp}.

\begin{figure}[h!]
	\begin{subfigure}{\textwidth}
		\centering
		\includegraphics[scale=0.5]{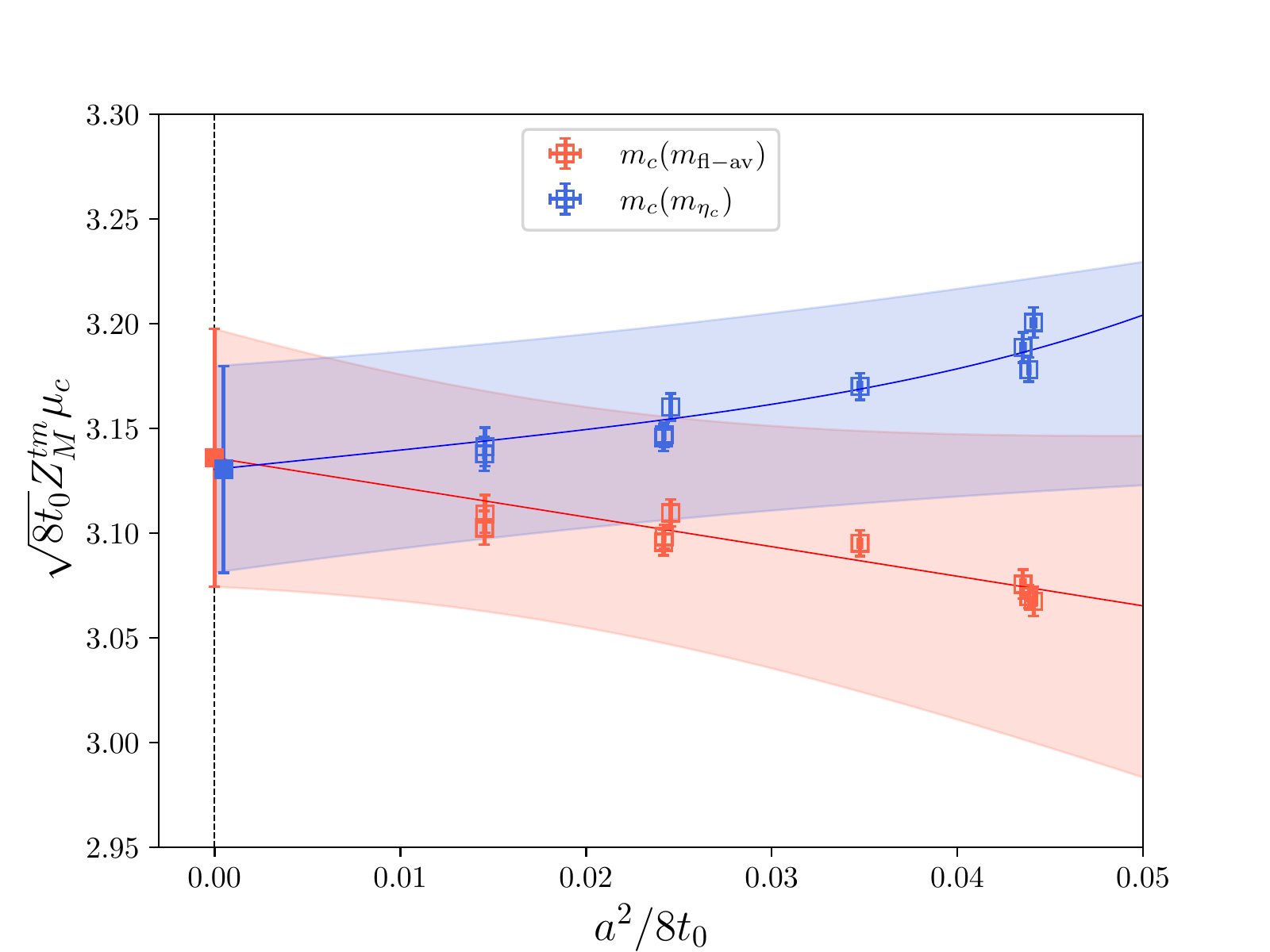} 
		\caption{Lattice spacing dependence of the renormalised charm quark mass for two representative model categories. The red points represent the charm quark mass fixed with the flavour average $m_{\bar{D}}$, while the blue points show results for the $\eta_c$ mass matching procedure. }
	\end{subfigure}
	\\
	\begin{subfigure}{0.5\textwidth}
		\includegraphics[scale=0.45]{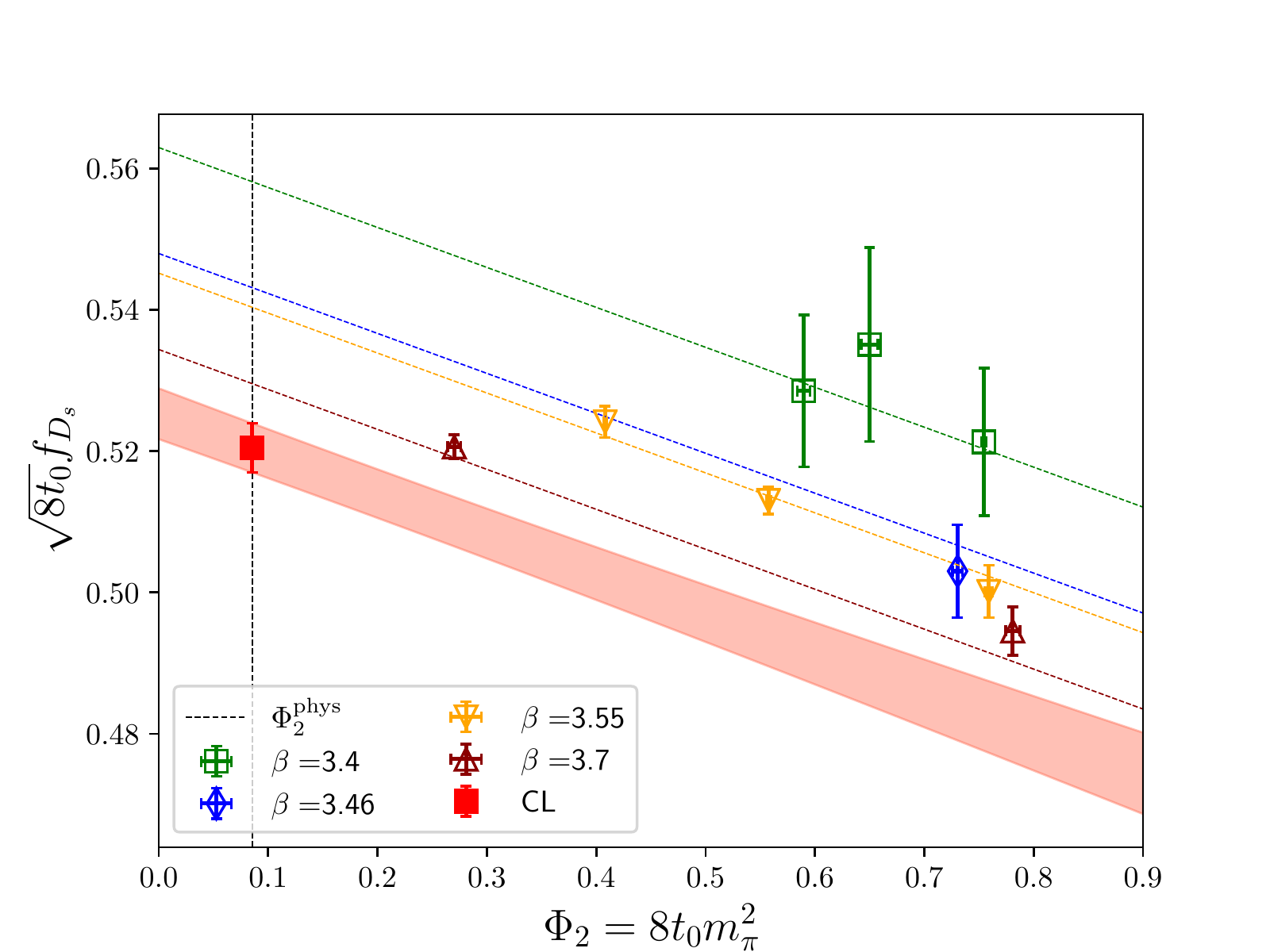}
		\caption{Pion mass dependence of charm-strange decay \\ constant $f_{D_s}$ for a representative category. }
	\end{subfigure}
	\begin{subfigure}{0.5\textwidth}
		\includegraphics[scale=0.45]{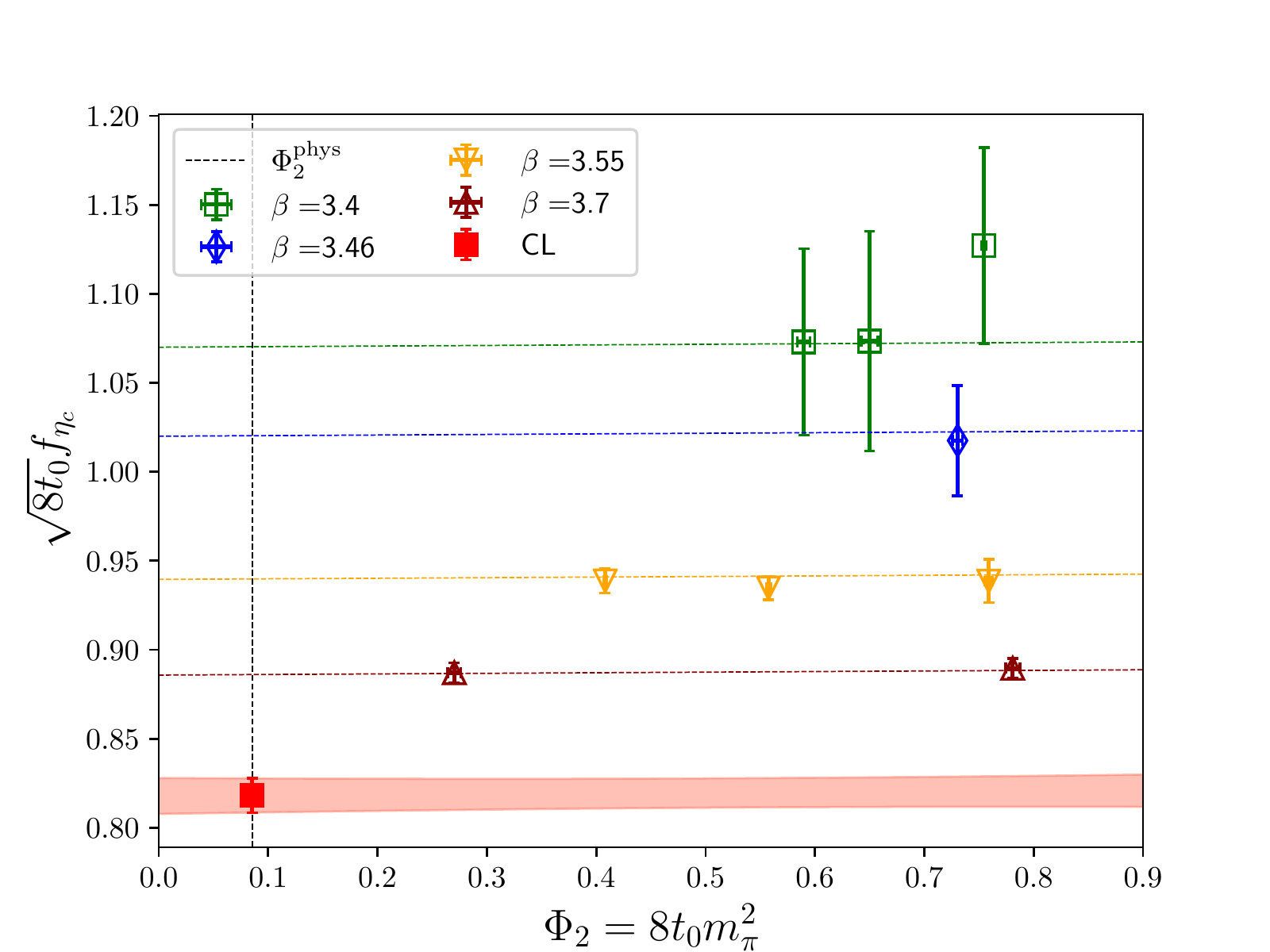}
		\caption{Pion mass dependence of charmonium decay \\ constant $f_{\eta_c}$ for a representative category. }
	\end{subfigure}
	\caption{Comparison and examples of some of the best fits according to the AIC procedure. The two-dimensional representations are obtained  by projecting to their physical values all the dependencies other than the ones showed in the plots. The shaded bands in the plots at the bottom show the continuum limit results. }
	\label{fig:best_fit}
\end{figure}

  In order to estimate the systematic effects arising from the model selection, we further classify our data into four different categories:
   we fix the physical charm mass by either employing the flavour averaged $D$ meson mass or the charmonium ground state $m_{\eta_c}$ and 
  	we either include all the gauge configurations in Tab. \ref{tab:ens} or exclude the ones with $\beta = 3.4$ to  analyze the impact of cutoff effects when removing the coarsest lattice spacing.
Within each category we perform fits by taking into account all the  possible models arising from Eq. (\ref{eq:tot_model}). This results in a total of 256 models for each observable. In Fig. \ref{fig:best_fit} we show examples of some of the best chiral-continuum extrapolations for different categories.

\begin{figure}[h!]
	
	\begin{subfigure}{0.5\textwidth}
		\centering
		\includegraphics[width=9cm,height=5cm]{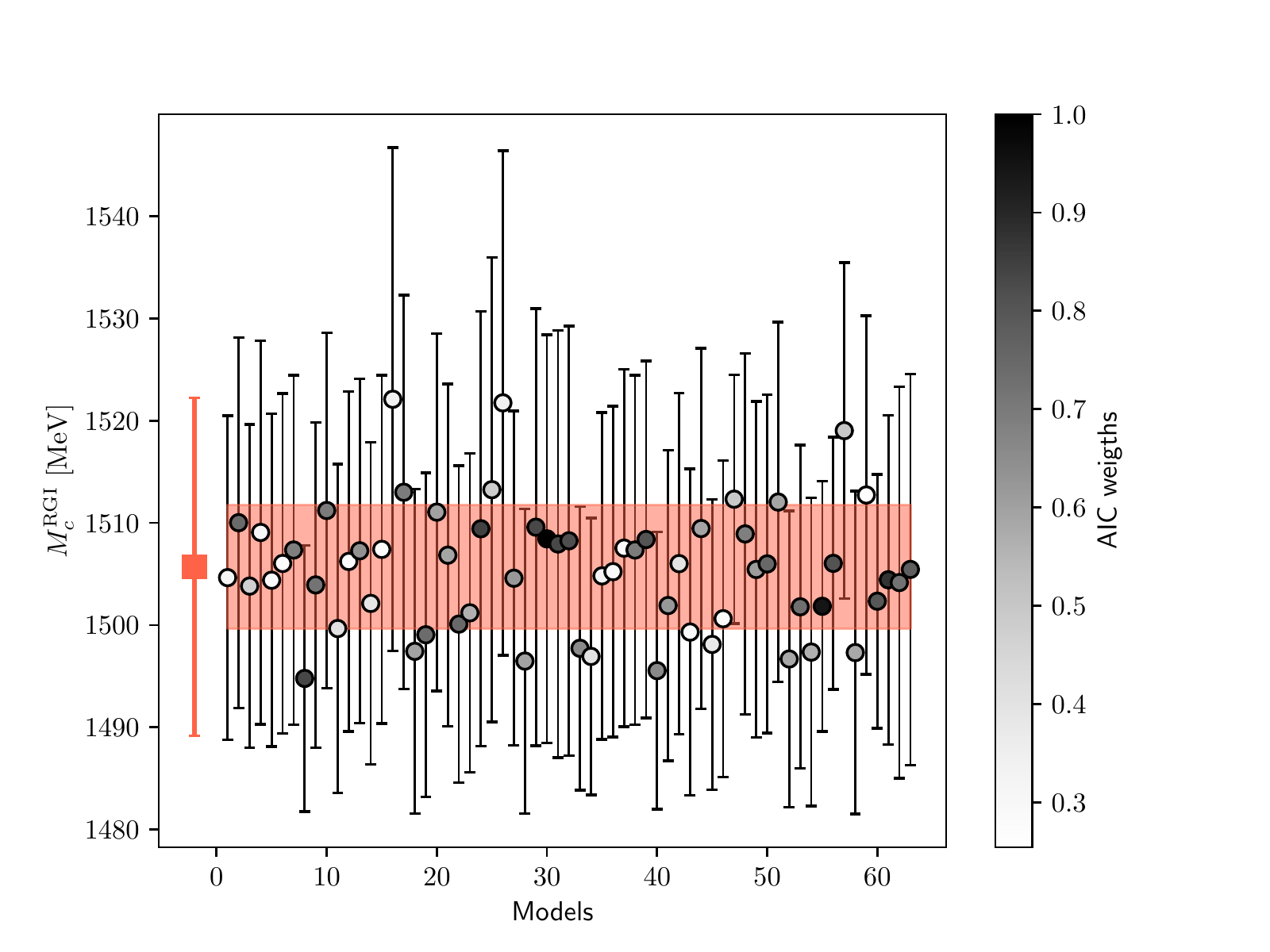}
	\end{subfigure}
	\begin{subfigure}{0.5\textwidth}
		\centering
		\includegraphics[width=6cm,height=5cm]{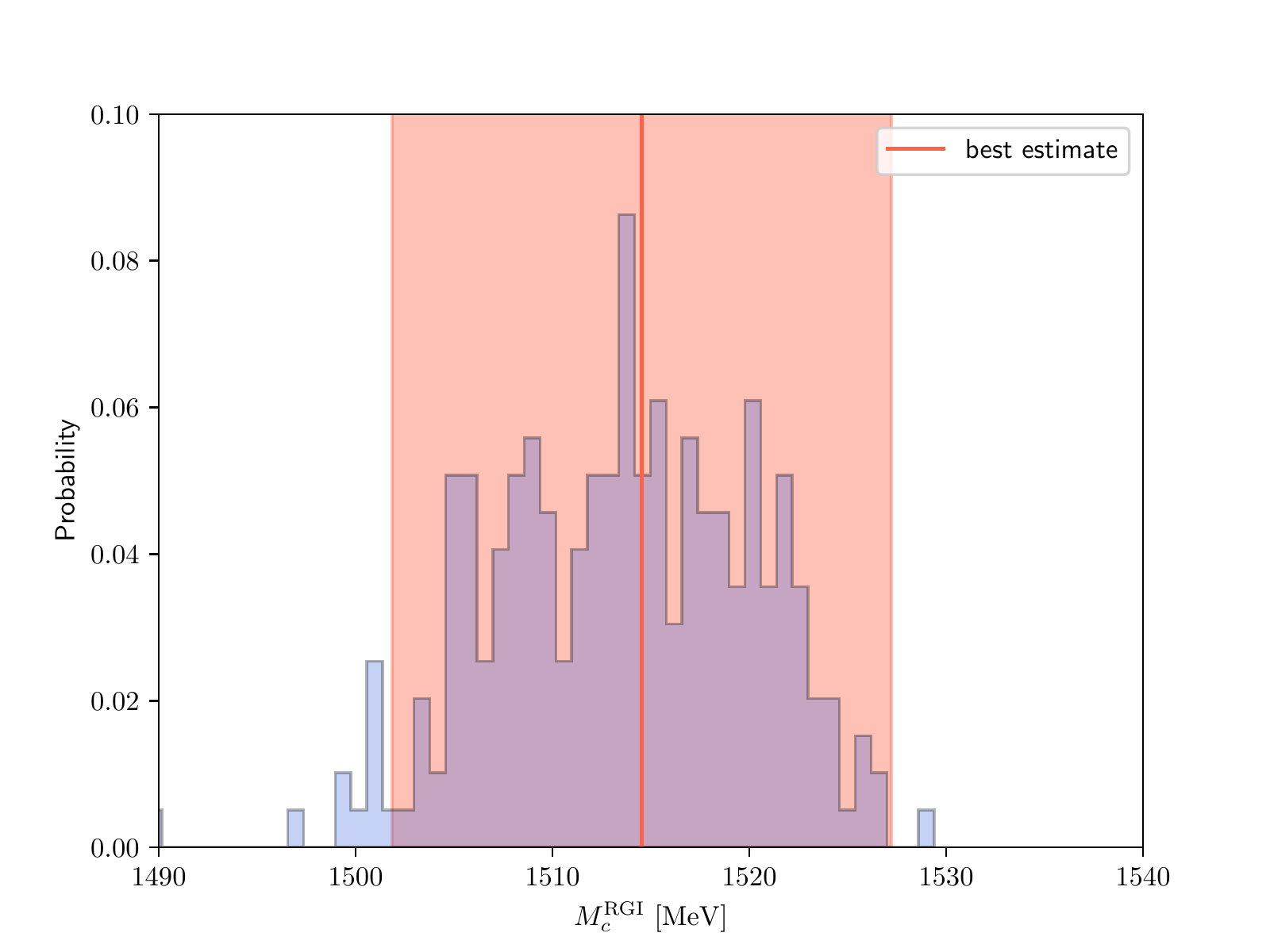}
	\end{subfigure}
	\caption{Summary of the model average procedure for the renormalised charm quark mass. Left: each circle labels  the fit result for a given model in Eq. (\ref{eq:tot_model}) for a representative category. The opacity of each circle is associated with the corresponding  AIC weight in the model average. The filled box labels the model average as extracted according to Eq. (\ref{eq:model_av}). Right: histogram with fit results of each model in each category. In both plots the red shaded bands label the systematic error according to Eq. (\ref{eq:syst}).  }
	\label{fig:best_cat}
\end{figure}

In order to classify the goodness of each fit in a given category we use the Akaike Information Criteria (AIC) in the same way as employed in  \cite{charm_quark}. To this purpose we introduce the AIC parameter as $\text{AIC} = (N-k) \frac{\chi^2}{\chi^2_{\text{exp}}} +2k$,
where $N$ is the number of data points and $k$ the number of parameters in the fit function.
Then we proceed with a model average within each category according to 
\begin{equation}\label{eq:model_av}
	\langle \mathcal{O} \rangle = \sum_{m=1}^{M}w_m \langle \mathcal{O}\rangle_m \qquad \text{where} \qquad 	w_m = \exp(-\frac{1}{2}\text{AIC}_m) \quad \text{s.t.} \quad \sum_{m=1}^{M}w_m = 1,
\end{equation}
where $\mathcal{O}$ is a given observable, $M$ labels the number of fit models, while $w_m$ are the corresponding weights given by the AIC.
As proposed in \cite{charm_quark}, we  estimate the  systematic error arising from the model selection  via 
\begin{equation}\label{eq:syst}
	\sigma_\mathcal{O}^2 = \sum_{m=1}^{M} w_m\langle\mathcal{O}\rangle_m^2 - \bigg(
	\sum_{m=1}^{M}w_m\langle\mathcal{O}\rangle_m
	\bigg)^2.
\end{equation}
In Fig. \ref{fig:best_cat} we report a graphical representation of the model average procedure.

\section{Preliminary results}
\noindent
 In order to extract our preliminary results we choose the weighted average over the best results on all categories. We employ again Eq. (\ref{eq:model_av}), using as weights the inverse of the systematic and statistical errors added in quadrature. The final systematic error is estimated from the weighted standard deviation in Eq. (\ref{eq:syst}). Following the chiral-continuum extrapolation analysis described in the previous section, we quote the ensuing results for charm quark observables in the $N_f=2+1$ theory. Here the first error is statistical and the second systematic.
 \begin{itemize}
 	\item RGI charm quark mass:  	$\ M_c(N_f=3) = 1514(13)(3)\ \text{MeV}$
 	
 \item D-mesons decay constant: 
 $\	f_D = 209(3.5)(1.5)\ \text{MeV} \qquad f_{D_s}= 242(4)(1)\ \text{MeV}$
\item Charmonium decay constants:\footnote{Neglecting disconnected contributions for $f_{\eta_c}$.}
$\ f_{\eta_c} = 392(8)(9)\ \text{MeV}\qquad f_{J/\Psi}= 395(10)(4)\ \text{MeV}$
 \end{itemize}

\begin{figure}[h!]
	
	\begin{subfigure}{0.5\textwidth}
		\centering
		\includegraphics[scale=0.4]{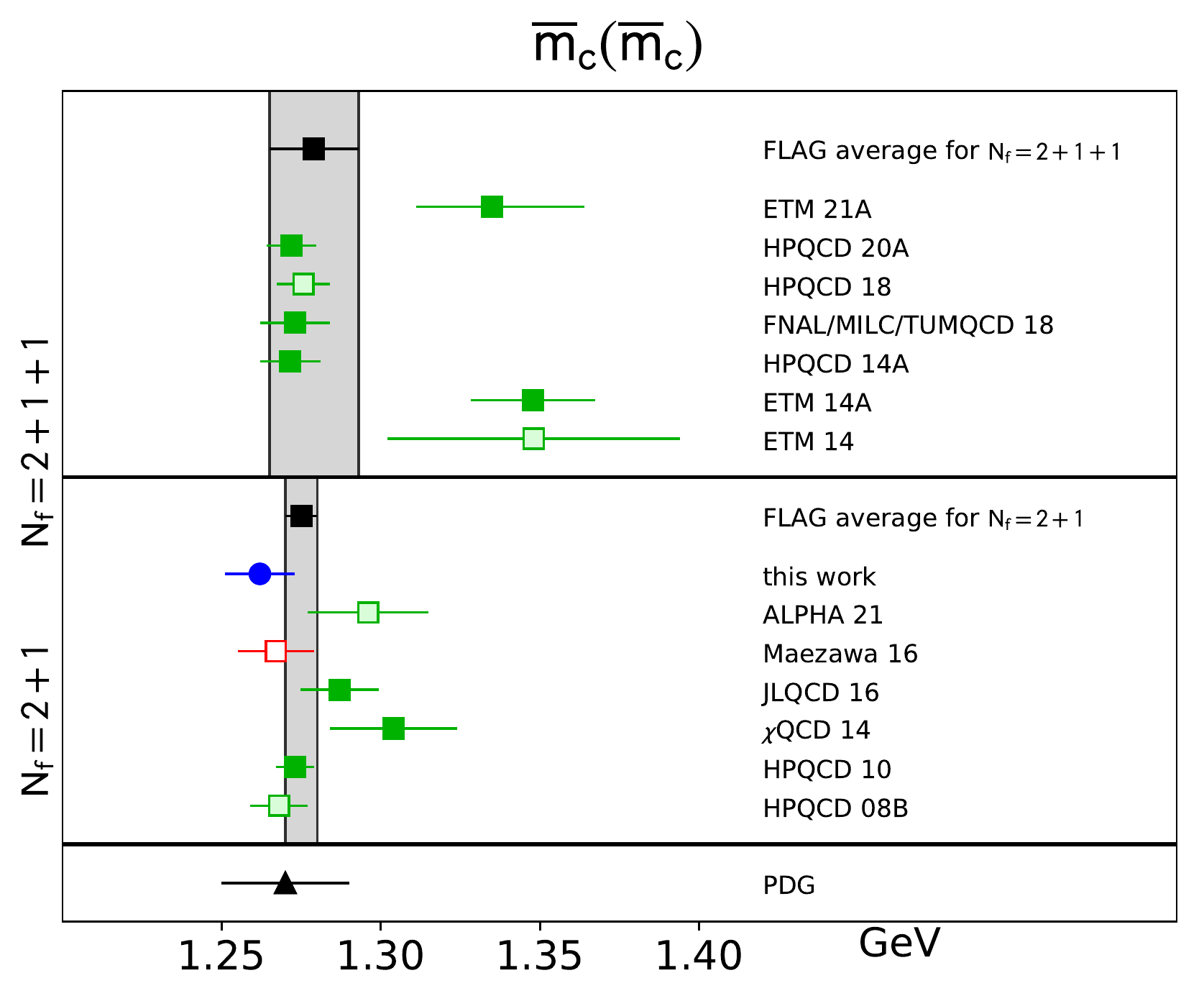}
	\end{subfigure}
	\begin{subfigure}{0.5\textwidth}
		\centering
		\includegraphics[scale=0.4]{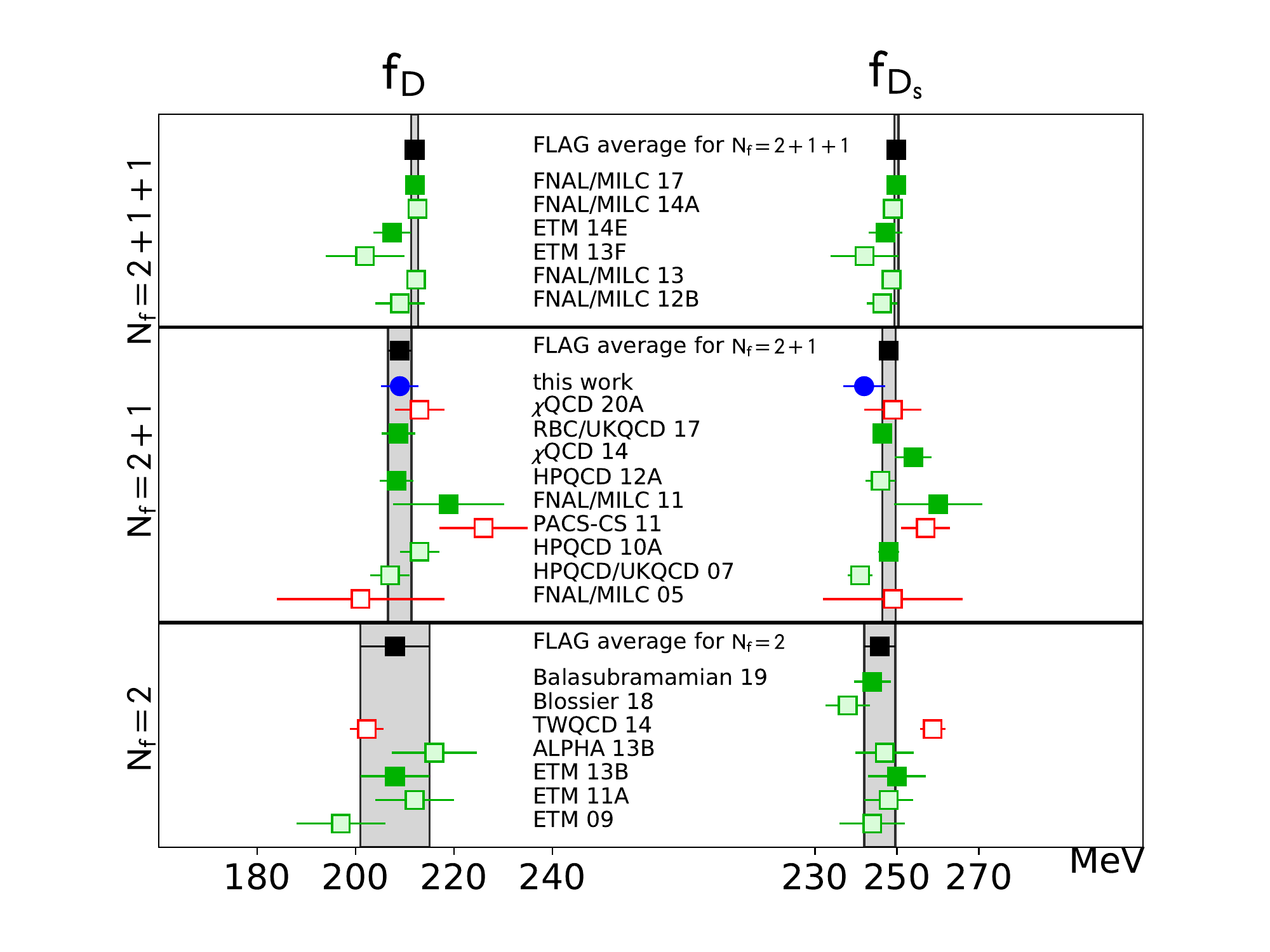}
	\end{subfigure}
	\caption{FLAG-style summary \cite{FLAG} of existing lattice results for the $\overline{\text{MS}}$-charm quark mass $\bar{m}_c(\bar{m}_c)$ in GeV \cite{mc} (left) together with decay constants of the $D$ and $D_s$ mesons in MeV \cite{fd} (right). The blue circles label our results. }
	\label{fig:fin_res}
\end{figure}

 \section{Conclusion and outlook}
\noindent
We have presented preliminary results from a tmQCD mixed-action setup at full twist for charm-light and charmonium decay constants and the RGI charm quark mass on a subset of CLS $N_f=2+1$ ensembles. In particular we highlight the usage of a GEVP variational method and an ongoing detailed analysis of the systematics in the chiral-continuum extrapolations. In order to improve the determination of charm-like observables a complete analysis including  the most chiral and the finest CLS ensembles will be considered in the following stage of the project.

\acknowledgments
\noindent
We acknowledge PRACE and RES for giving us access to computational resources at MareNostrum (BSC). We thank CESGA for granting access to Finis Terrae II. This work is supported by the European Union's Horizon 2020 research and innovation programme under grant agreement No 813942 and by the Spanish MINECO through project PGC2018-094857-B-I00, the Centro de Excelencia Severo Ochoa Programme through SEV-2016-0597
and the Ramón y Cajal Programme RYC-2012-0249. We are grateful to CLS members for producing the gauge configuration ensembles used in this study.

\end{document}